\documentclass[12pt]{article}
\usepackage{epsfig}
\usepackage{multirow}
\usepackage{cite}
\usepackage{graphicx}
\usepackage{amssymb}
\usepackage{amsmath}
\usepackage{latexsym}

\newcommand{\be}{\begin{eqnarray}}
\newcommand{\ee}{\end{eqnarray}}
\newcommand{\ra}{\rightarrow}

\def\norm@note#1#2{\special{}
  \ifinner{\ifdim\baselineskip=\z@
    \baselineskip18\p@\fi
    \ifhmode
    \raisebox{.5\baselineskip}[\z@][\z@]{%
      \rlap{\sf\scriptsize #2}}%
    \else\vskip-\baselineskip%
    \raisebox{-.6\baselineskip}[\z@][\z@]{%
      \rlap{\sf\scriptsize #2}}%
    \fi}%
  \else\marginpar{\raggedright\if@twoside\ifodd\c@page%
    \raggedleft\fi\fi\sf\scriptsize #1#2}%
  \fi}%

\begin{document}

\begin{titlepage}

\vskip.40cm
\begin{center}
{\Large \bf Measuring $Z'$ couplings at the LHC} \vskip.5cm
\end{center}
\vskip0.2cm

\begin{center}
{\bf Frank Petriello and Seth Quackenbush}
\end{center}
\vskip 8pt
\begin{center}
{\it Department of Physics, University of Wisconsin, Madison, WI 53706, USA} \\
\end{center}

\vglue 0.3truecm

\begin{abstract}
\vskip 3pt \noindent

We study the properties of potential new $Z'$ gauge bosons produced through the Drell-Yan mechanism at the LHC.  Our analysis is performed using a fully differential 
next-to-leading order QCD calculation with spin correlations, interference effects, and experimental acceptances included.  We examine the distinguishability of different models and the feasibility of extracting general coupling information with statistical, residual scale, and current parton distribution function error estimates included.  We extend a  
previous parametrization of $Z'$ couplings to include parity-violating coupling combinations, and introduce a convenient technique for simulating new gauge bosons on-peak using the concept of basis models.  We illustrate our procedure using several example $Z'$ models.  We find that one can extract reliably four combinations of generation-independent quark and lepton couplings in our analysis.  For a $Z'$ mass of 1.5 TeV, one can determine coupling information very well assuming $100 \, {\rm fb}^{-1}$ of integrated luminosity, and a precise measurement becomes possible with $1 \, {\rm ab}^{-1}$ at the SLHC.  For  a 3 TeV mass, a reasonable determination requires the SLHC.

\end{abstract}

\end{titlepage}
\newpage
\section{Introduction \label{intro}}

$Z'$ gauge bosons that appear in $U(1)$ gauge extensions of the Standard Model (SM) are the most 
ubiquitous particles in models of new physics.  They appear in grand unified theories such as $SO(10)$~\cite{mohapatra} and $E(6)$~\cite{Hewett:1988xc}, in Little Higgs models~\cite{Schmaltz:2005ky}, and in theories with extra space-time dimensions~\cite{Hewett:2002hv}.  They often appear as messengers 
which connect the SM to hidden sectors, such as in some models of supersymmetry breaking~\cite{Chung:2003fi} and in Hidden Valley 
models~\cite{Strassler:2006im}.  $Z'$ states that decay to lepton pairs have a simple, clean experimental signature and can easily be searched for at 
colliders.  Current direct search limits from the Tevatron require the $Z'$ mass to be greater than about 900 GeV when its couplings to SM fermions are identical to those of the $Z$ boson~\cite{Tev:2007sb}.

Since the experimental signature is clean and the QCD uncertainties for inclusive quantities such as the total cross section and $p_T$ 
spectrum have been studied and found to be fairly small~\cite{Fuks:2007gk}, it is likely that the couplings of a discovered $Z'$ can be studied with reasonable 
accuracy to probe the high scale theory that gave rise to it.  Many studies of how to measure $Z'$ properties and couplings to SM particles have 
been performed~\cite{zprevs}.  In particular, a recent study focusing on Tevatron physics introduced a parametrization of the parity symmetric 
couplings of the $Z'$ to SM fermions that allows for a simple comparison between experimental measurements and theoretical models~\cite{Carena:2004xs}.  

We attempt to extend previous studies of $Z'$ coupling extractions in several ways in this paper.
\begin{itemize}

\item We perform a next-to-leading order (NLO) QCD calculation of the fully differential $pp \to (\gamma,Z,Z') \to l^+l^- X$ cross section with all interference 
effects and spin correlations included.  We also include realistic LHC acceptance cuts.

\item We study the effect of statistical, parton distribution function (PDF), and residual scale errors on important $Z'$ observables such as the 
total cross section, forward-backward asymmetries on and off-peak, and central to forward rapidity ratio.

\item We extend the parametrization of $Z'$ couplings in~\cite{Carena:2004xs} to include parity-violating coupling combinations, which can 
be accessed when differential measurements are made.

\item We introduce the use of basis models when simulating $Z'$ states.  These arise from the observation that the differential cross section 
can be written as a product of $Z'$ couplings multiplied by functions that depend significantly on the specific $Z'$ under consideration only through its mass.  
These functions depend on the PDFs, matrix elements, acceptance cuts, and details of the experimental analysis, but need be computed only once 
for a given $Z'$ mass.  They can be obtained by running $Z'$ simulation codes for basis vectors in coupling space.  This facilitates the 
separation of the specifics of a given $Z'$ model from the details of QCD and the experimental analysis.  The functions we introduce encoding these details 
are extensions of the $w_{u,d}$ introduced in~\cite{Carena:2004xs}.

\item We study the extraction of the parity symmetric and parity violating $Z'$ couplings at both the LHC and SLHC, and quantify the effect of 
statistical, PDF, and residual scale errors on the accuracy of their determination.  We note interesting correlations between these errors.  Using several example $Z'$ models, we examine how well the LHC and SLHC can distinguish between different models.

\end{itemize}
We use four example $Z'$ models to illustrate our techniques: three models arising from an $E(6)$ unified theory and one from a left-right symmetric model.  We find 
that residual QCD scale uncertainties have a negligible effect on the measurement of $Z'$ couplings.  Statistical and current PDF errors are larger and have 
an approximately equivalent effect at the LHC.  PDFs should become more accurately known improve when LHC data is used to constrain them.  We find that 
the parity symmetric couplings of a 1.5 TeV $Z'$ should be measured with good precision at the LHC.  Although some information can be obtained about the parity violating 
couplings at the LHC, a more accurate determination requires super-LHC (SLHC) statistics.  Measurements of the couplings of a 3 TeV $Z'$  require SLHC statistics.

Our paper is organized as follows.  In Section~\ref{tframe} we review the theoretical assumptions underlying our analysis and present details of the four example $Z'$ models we consider.  In Section~\ref{bobs} we 
present the $Z'$ observables used in our analysis.  We discuss the details of our calculation and present results showing the effect of statistical, PDF, and residual scale uncertainties on the basic $Z'$ observables in Section~\ref{calcframe}.  The bulk of our study is discussed in Section~\ref{cmeas}.  We introduce our
parity symmetric and parity violating $Z'$ coupling combinations, and show how to extract them from on-peak measurements using the concept of 
basis models.  We also study how well the LHC and SLHC can determine $Z'$ couplings using our four example models for illustration.  In Section~\ref{conc} we present 
our conclusions.

\section{Theoretical framework \label{tframe}}

We first describe the types of $Z'$ models we consider in our study.  We assume that the $Z'$ couplings are generation independent to avoid large flavor changing 
neutral currents that would restrict $M_{Z'}$ to be 100 TeV or more.  We also assume that the generator of the $U(1)$ group giving rise to the $Z'$ state commutes with 
the Standard Model $SU(2)_L$ generators.  This implies that the couplings of the $Z'$ to $u_L$ and $d_L$, the members of an $SU(2)_L$ quark doublet $q_L$, are the same, 
as are the couplings of the $Z'$ to the members of a lepton doublet $l_L$.  These restrictions leave us with the following five parameters: the coupling to $q_L$, the coupling to $l_L$, and the three couplings to the $SU(2)_L$ singlet states $u_R$, $d_R$, and $e_R$.  We have absorbed the overall gauge coupling into these five quantities.  We neglect possible 
$Z-Z'$ mixing; LEP $Z$-pole measurements restrict the mass mixing angle to be less than approximately $10^{-3}$ radians~\cite{Abreu:1995}. 

We utilize four models representative of $Z'$ models discussed in the literature as examples to illustrate the extraction of $Z'$ couplings from LHC data.  In particular, we examine three possible $U(1) \, Z'$ bosons originating from the exceptional group $E_{6}$, and one coming from a left-right symmetric model, which can arise from an $SO(10)$ GUT.  We describe these models below and list the couplings of the SM 
fermions to the $Z'$.

\begin{itemize}

\item \underline{$E_{6}$}: $E_{6}$ models are described by the breaking chain
\be
E_{6} \ra SO(10) \times U(1)_{\psi} \ra SU(5) \times U(1)_{\chi} \times U(1)_{\psi} \ra SM \times U(1)_{\beta}
\ee
where
\be
Z' = Z'_{\chi} \cos\beta + Z'_{\psi} \sin\beta
\ee
is the lightest new boson arising from this breaking.  In this paper we examine the $\chi$ model ($\beta = 0$), the $\psi$ model ($\beta = \pi/2$), and the $\eta$ model ($\beta = \arctan(-\sqrt{5/3})$).

\item \underline{Left-right symmetric models}: We also consider a left-right model coming from the symmetry group $SU(2)_{R} \times SU(2)_{L} \times U(1)_{B-L}$.  Left-right models can arise from the following breaking of $SO(10)$:
\be
SO(10) \ra SU(3) \times SU(2)_{L} \times SU(2)_{R} \times U(1)_{B-L} .
\ee
The $Z'$ in left-right models couples to the current
\be
J^{\mu}_{LR} = \alpha_{LR} J^{\mu}_{3R} - \frac{1}{2\alpha_{LR}} J^{\mu}_{B-L}, 
\ee
with $\alpha_{LR} = \sqrt{(c^{2}_{W}g^{2}_{R}/s^{2}_{W}g^{2}_{L}) - 1}$, and $g_{L} = e/\sin\theta_{W}$.  In the symmetric case $g_{L} = g_{R}$ and $\alpha_{LR} \simeq 1.59$, using the on-shell definition of $\sin^2 \theta_W$.  The overall coupling strength is $e/\cos\theta_W$.  We use the symmetric model in our analysis.

\end{itemize}

The fermion coupling assignments in these models are summarized in the following table.  For convenience, we factor out an overall $e/\cos\theta_{W}$.  We assume that the couplings for $E_{6}$ models retain their GUT-scale relations to the EM coupling down to the $Z'$ scale to good approximation.  Assuming a different value for the overall coupling may have some impact on the total cross section.  However, the statistical uncertainties for each model does not vary significantly if the overall coupling is allowed to vary 20-30\%.

\begin{table}[h]
\begin{center}
\begin{tabular}{| c || c | c | c | c |}
\hline
 & $\chi$ & $\psi$ & $\eta$ & $LR$ \\ \hline \hline
 $q_{L}$ & $\frac{-1}{2\sqrt{6}}$ & $\frac{\sqrt{10}}{12}$ & $1/3$ & $\frac{-1}{6\alpha_{LR}}$ \\ \hline
 $u_{R}$ & $\frac{1}{2\sqrt{6}}$ & $\frac{-\sqrt{10}}{12}$ & $-1/3$ & $\frac{-1}{6\alpha_{LR}}+\frac{\alpha_{LR}}{2}$ \\ \hline
 $d_{R}$ & $\frac{-3}{2\sqrt{6}}$ & $\frac{-\sqrt{10}}{12}$ & $1/6$ & $\frac{-1}{6\alpha_{LR}}-\frac{\alpha_{LR}}{2}$ \\ \hline
 $l_{L}$ & $\frac{3}{2\sqrt{6}}$ & $\frac{\sqrt{10}}{12}$ & $-1/6$ & $\frac{1}{2\alpha_{LR}}$ \\ \hline
 $e_{R}$ & $\frac{1}{2\sqrt{6}}$ & $\frac{-\sqrt{10}}{12}$ & $-1/3$ & $\frac{1}{2\alpha_{LR}}-\frac{\alpha_{LR}}{2}$ \\ \hline
\end{tabular} 
\end{center}
\caption{Fermion couplings to the $Z'$ for the considered models.  An overall $e/\cos\theta_{W}$ has been factored out.}
\end{table}

\section{Basic observables \label{bobs}}

If a $Z'$ is discovered at the LHC, the next step will be to determine the underlying model from which it arises.  The following observables can be used to check whether the 
data fits the hypothesis of a certain model and begin to measure $Z'$ properties.

\begin{itemize}
\item $Z'$ mass and total width, $M_{Z'}$ and $\Gamma_{Z'}$.

One should be able to find a peak at the LHC from an excess of dilepton events.  The location of the resonance determines the mass of the $Z'$.  The width is determined 
by fitting the resonance peak to the Breit-Wigner form $1/\left[(M_{ll}^2-M_{Z'}^2)^2+M_{Z'}^2\Gamma_{Z'}^2\right]$.  The width is sensitive to $Z'$ couplings to all 
final states, and can probe invisible decay modes.  We assume that the $Z'$ has no invisible decays besides to neutrinos in our example models.  Our analysis is nearly width-independent, and our results will not differ provided the invisible width is not large enough to make the $Z'$ too broad.

\item Cross section to $e^{+} e^{-}$, $\sigma$.  

In defining the $Z'$ on-peak cross section, we follow~\cite{Dittmar:2003ir} and keep events within $\pm 3 \Gamma$ of the resonance peak.  Using this instead of a fixed value allows for more consistency between models in isolating the $Z'$ from the other neutral gauge bosons, and renders the cross section less sensitive to the width chosen.  A nearly width-independent quantity is $\sigma \Gamma$.  The total cross section provides a good first separation between models and gives an indication of overall coupling strength and leptonic branching fraction.

\item Forward-backward asymmetry, $A_{FB}$.

$A_{FB}$ measures the relative difference of forward-scattered events and backward-scattered events:

\be
A_{FB}^{y_{1}} = \frac{[\int_{y_{1}}^{y_{max}} - \int_{-y_{max}}^{-y_{1}}] [F(y) - B(y)] dy}{[\int_{y_{1}}^{y_{max}} + \int_{-y_{max}}^{-y_{1}}] [F(y) + B(y)] dy}
\ee
where $F(y) = \int_{0}^{1} d\cos\theta \frac{d^{2}\sigma}{dy d\cos\theta}$, $B(y) = \int_{-1}^{0} d\cos\theta \frac{d^{2}\sigma}{dy d\cos\theta}$, $y$ is the $Z'$ rapidity, and 
$y_{max}$ is the maximum allowed $Z'$ rapidity given by $\frac{1}{2}  {\rm ln} (s/M_{Z'}^2)$.  The electron-quark angle $\theta$ is taken to be in the Collins-Soper 
frame~\cite{Collins:1977iv}, but there is an ambiguity in the quark vs. anti-quark direction, since it is unknown which proton carried it.  We follow the suggestion in~\cite{Dittmar:1996my} and choose the quark direction along the direction of the $Z'$ rapidity.  Equivalently, one could choose one beam as the quark direction and and exploit the antisymmetry in $y$~\cite{Langacker:1984dc}.  The value of $y_{1}$ can be chosen to throw away events with low $Z'$ rapidity, i.e., those where the quark direction is more likely to be misidentified.  We study the dependence on $y_{1}$ in our analysis.  $A_{FB}$ is quite sensitive to models with parity violating couplings.  The on-peak 
value of $A_{FB}$ is defined by keeping events within $\pm 3 \Gamma$ of the resonance peak, as for the cross section.


\item Rapidity ratio, $R$.

The central/forward rapidity ratio is defined as

\be
R_{y_{1}} = \frac{\int_{-y_{1}}^{y_{1}}[F(y) + B(y)] dy}{[\int_{y_{1}}^{y_{max}} + \int_{-y_{max}}^{-y_{1}}] [F(y) + B(y)] dy}.
\ee
$R$ measures the ratio of central rapidities to extreme rapidities.  Since the up and down PDF distributions have substantially different profiles, they should weight $Z'$ events differently in rapidity, and thus $R$ can help distinguish up versus down couplings.  $R$ is defined by  keeping events within $\pm 3 \Gamma$ of the resonance peak.

\item Off-peak asymmetry, $A_{FB}^{off-peak}$.

In addition to the above on-peak observables, the profile of $A_{FB}$ in dilepton invariant mass bins below $M_{Z'}$ may vary considerably among models.  For this observable, we integrate instead in the region $2/3 M_{Z'} < M_{ll} < M_{Z'} - 3 \Gamma_{Z'}$.

\end{itemize}

\section{Calculation \label{calcframe}}

To study the $Z'$ signal at the LHC, we perform a fully differential next-to-leading-order QCD calculation of all observables considered.  We include all spin correlations and interferences between the 
photon, $Z$, and $Z'$.  We impose the following basic acceptance cuts on the final state lepton transverse momenta and pseudorapidities: $p_T^l > 20$ GeV and 
$|\eta^l| < 2.5$.  Previous studies have found that detector resolution effects and other measurement errors are unlikely to have a significant effect on the 
$e^+e^-$ final state~\cite{Dittmar:2003ir}, and are neglected.  A CMS simulation of $Z'$ production found reconstruction efficiencies near 90\% in the electron channel and no significant detector systematic errors~\cite{Clerbaux}.  In addition, electron energies can be measured to better than 1\% accuracy, and invariant masses can therefore be reconstructed very well.  The factorization and renormalization scales are taken to be $\mu_F=\mu_R= M_{Z'}$.  They are varied 
simultaneously from 
$M_{Z'}/2$ to $2 M_{Z'}$ to determine scale errors.  We use the CTEQ 6.5 NLO PDF sets \cite{Pumplin:2002vw}.  PDF error estimates are determined by calculating 
each observable with 
each of the 40 PDF eigenvector sets (corresponding to 20 +/- directions in parameter space), and combining the errors in each +/- shift in quadrature.  
Statistical errors are those that can be expected for an integrated luminosity of $100\, {\rm fb}^{-1}$ unless stated otherwise.  

We present below in Tables~\ref{basic_results},~\ref{asym}, and~\ref{results_0.4} for orientation the cross section, cross section times width, acceptance, asymmetries, and central-forward ratio for our four example models.  
We examine $Z'$ states with $M_{Z'} = 1.5$ TeV and $3$ TeV.  The acceptance denotes the fraction of events that pass the cuts on $p_T^l$ and $\eta^l$ presented above; the fully inclusive results are obtained by dividing the cross section results by the acceptances.  There are a few interesting features to note in these numbers.

\begin{itemize}

\item The acceptances after imposing realistic cuts are independent of the model considered to the percent level.  This was also observed in a simplified analysis 
in~\cite{Carena:2004xs}.

\item The acceptance, which is formed from a ratio of the cross section including cuts over the inclusive cross section, has tiny residual scale errors, 
indicating that it is insensitive to uncalculated higher order QCD corrections.  We have checked that the leading-order acceptance is nearly indentical to the NLO 
result.  It is generically true that scale errors are negligible for ratios of sufficiently inclusive quantities, and we see in Tables~\ref{asym} and~\ref{results_0.4} that it is true for the other quantities considered here such as $R$, $A_{FB}$, and $A_{FB}^{off-peak}$.  This indicates that NNLO QCD corrections have no effect on the analysis of $Z'$ properties.

\item While $y_1 = 0.8$ is the canonical choice for $A_{FB}$ measurements, this is an extreme rapidity for a $3$ TeV $Z'$ at the LHC.  Therefore, we also give $3$ TeV values for $y_1 = 0.4$ in Table~\ref{results_0.4}.

\item PDF errors are relatively large for the total cross section, and are not negligible for other observables.

\item NLO results are substantially different from LO, especially for the total cross section.  For instance, at LO, for the $\chi$ model at $1.5$ TeV, one finds $\sigma = 40.0$ fb $\pm^{2.5}_{2.3}$ and $A_{FB}^{0.8} = -0.2137 \pm^{0.0003}_{0.0002}$ (errors are scale) versus $50.3$ fb and $-0.217$ at NLO.  NLO corrections are larger than LO scale errors would suggest.  This is not surprising; similar results have been seen for $Z$ production for the scale range considered here~\cite{DYrefs}.  NNLO corrections to the $Z$ cross section have been shown to leave the NLO central value essentially unchanged while further reducing the scale error, and so we have confidence in our NLO analysis.
\end{itemize}

Before continuing, we note a caveat regarding our calculation.  We have not included higher-order electroweak effects.  The complete ${\cal O}(\alpha)$ 
electroweak corrections for SM $l^+l^-$ production for hadron colliders have been calculated in~\cite{Baur:2001ze}.  Several components of the higher-order electroweak 
corrections can be identified, and we indicate below how we expect them to affect our study.

\begin{itemize}

\item QED corrections can be separated into two classes, those associated with initial and final state radiation.  Those associated with initial state 
radiation lead to collinear singularities that must be absorbed into the bare PDFs, and modify the DGLAP evolution of the PDFs.  The initial-state QED 
corrections in a DIS scheme were shown to be at the percent level or smaller~\cite{Baur:2001ze}.  This study did not include the 
QED effects on PDF evolution, as the appropriate PDF sets were unavailable at the time.  The initial-state QED effects on DGLAP evolution have since been 
incorporated into a global fit to the available data~\cite{Martin:2004dh}, and they appear unlikely to have a significant affect on our analysis.

\item Final state QED radiation can have a significant effect on the lepton pair invariant mass distribution~\cite{Baur:1997wa,Baur:2001ze}.  
They should be included in a more complete analysis.

\item Weak corrections consist of factorizable terms that can be absorbed into effective couplings and masses, and non-factorizable pieces arising 
from box diagrams that cannot.  Since we treat the $Z'$ couplings to fermions as free parameters in our coupling extraction, the factorizable corrections 
have no effect on our study.  They enter only when we choose a value of $\sin^{2}\theta_{W}$ for the left-right model results.

\item The non-factorizable corrections arising from electroweak logarithms of the form $\ln(\frac{s}{M_{W,Z}^2})$ become large for high lepton pair invariant masses in the Standard Model.  While we expect these contributions to be small on the $Z'$ peak, they can become important in off-peak observables.  The new box diagrams are those containing both a $Z'$ and a $Z$.  These should contain only a single logarithm~\cite{Ciafaloni:1998xg}, 
and we expect them to be less important than for Standard Model lepton pair production at high energies.

\end{itemize}

\begin{table}[ht]
\begin{center}
\begin{tabular}{| c | c || c l | c l | c c |}
\hline
 $M_{Z'}$ & Model & \multicolumn{2}{|c|}{$\sigma$ (fb)} & \multicolumn{2}{|c|}{$\sigma \Gamma$ (fb $\cdot$ GeV)} & \multicolumn{2}{|c|}{Acceptance} \\ \hline \hline
 \multirow{12}{*}{1.5 TeV} & & & $\pm 0.71$ & & $\pm 12.7$ & & - \\
 & $\chi$ & $50.26$ & $\pm^{3.13}_{2.97}$ & $898.3$ & $\pm^{55.9}_{53}$ & $0.9067$ & $\pm^{0.0035}_{0.0044}$ \\
 & & & $\pm^{1.21}_{1.22}$ & & $\pm^{21.6}_{21.9}$ & & $\pm^{0.0002}_{0.0002}$ \\ \cline{2-8}

 & & & $\pm 0.49$ & & $\pm 4$ & & - \\
 & $\psi$ & $24.25$ & $\pm^{1.34}_{1.22}$ & $197.4$ & $\pm^{10.9}_{9.9}$ & $0.8957$ & $\pm^{0.0020}_{0.0022}$ \\
 & & & $\pm^{0.57}_{0.56}$ & & $\pm^{4.6}_{4.6}$ & & $\pm^{0.0005}_{0.0001}$ \\ \cline{2-8}

 & & & $\pm 0.53$ & & $\pm 5.1$ & & - \\
 & $\eta$ & $27.79$ & $\pm^{1.5}_{1.37}$ & $271.0$ & $\pm^{14.6}_{13.3}$ & $0.8928$ & $\pm^{0.0018}_{0.0023}$ \\
 & & & $\pm^{0.64}_{0.64}$ & & $\pm^{6.2}_{6.3}$ & & $\pm^{0.0003}_{0.0002}$ \\ \cline{2-8}

 & & & $\pm 0.78$ & & $\pm 26$ & & - \\
 & $LR$ & $60.71$ & $\pm^{3.44}_{3.19}$ & $2049$ & $\pm^{116}_{108}$ & $0.8995$ & $\pm^{0.0022}_{0.0029}$ \\
 & & & $\pm^{1.43}_{1.44}$ & & $\pm^{48}_{49}$ & & $\pm^{0.0002}_{0.0001}$ \\ \hline \hline

 \multirow{12}{*}{3 TeV} & & & $\pm 0.11$ & & $\pm 3.8$ & & - \\
 & $\chi$ & $1.12$ & $\pm^{0.12}_{0.11}$ & $40.5$ & $\pm^{4.4}_{3.9}$ & $0.9477$ & $\pm^{0.0021}_{0.0028}$ \\
 & & & $\pm^{0.04}_{0.05}$ & & $\pm^{1.5}_{1.6}$ & & $\pm^{0.0002}_{0.0001}$ \\ \cline{2-8}

 & & & $\pm 0.078$ & & $\pm 1.28$ & & - \\
 & $\psi$ & $0.608$ & $\pm^{0.05}_{0.045}$ & $9.97$ & $\pm^{0.81}_{0.74}$ & $0.9449$ & $\pm^{0.0010}_{0.0014}$ \\
 & & & $\pm^{0.022}_{0.023}$ & & $\pm^{0.35}_{0.38}$ & & $\pm^{0.0002}_{0.0003}$ \\ \cline{2-8}

 & & & $\pm 0.085$ & & $\pm 1.7$ & & - \\
 & $\eta$ & $0.716$ & $\pm^{0.057}_{0.052}$ & $14.1$ & $\pm^{1.1}_{1.0}$ & $0.9442$ & $\pm^{0.0012}_{0.0013}$ \\
 & & & $\pm^{0.025}_{0.027}$ & & $\pm^{0.5}_{0.5}$ & & $\pm^{0.0002}_{0.0002}$ \\ \cline{2-8}

 & & & $\pm 0.12$ & & $\pm 8.2$ & & - \\
 & $LR$ & $1.46$ & $\pm^{0.11}_{0.11}$ & $99.7$ & $\pm^{7.8}_{7.8}$ & $0.9457$ & $\pm^{0.0022}_{0.0029}$ \\
 & & & $\pm^{0.05}_{0.06}$ & & $\pm^{3.6}_{3.9}$ & & $\pm^{0.0002}_{0.0001}$ \\ \hline
\end{tabular}
\end{center}
\caption{Cross sections and acceptances for our example $Z'$ models.  From top to bottom, the errors are statistical ($100 \, {\rm fb}^{-1}$), PDF, and scale.  Errors in $\sigma \cdot \Gamma$ are those due to the cross section only.  As the acceptance is a theoretical quantity, we do not report a statistical error.}
\label{basic_results}
\end{table}

\begin{table}[ht]
\begin{center}
\begin{tabular}{| c | c || c l | c l | c l |}
\hline
 $M_{Z'}$ & Model & \multicolumn{2}{|c|}{$A_{FB}^{0.8}$} & \multicolumn{2}{|c|}{$A_{FB}^{off-peak,0.8}$} & \multicolumn{2}{|c|}{$R_{0.8}$} \\
\hline \hline
 \multirow{12}{*}{1.5 TeV} & & & $\pm 0.025$ & & $\pm 0.051$ & & $\pm 0.073$ \\
 & $\chi$ & $-0.217$ & $\pm^{0.019}_{0.016}$ & $0.164$ & $\pm^{0.007}_{0.008}$ & $2.362$ & $\pm^{0.138}_{0.155}$ \\
 & & & $\pm^{0.001}_{0}$ & & $\pm^{0.001}_{0.002}$ & & $\pm^{0.009}_{0.008}$ \\ \cline{2-8}

 & & & $\pm 0.035$ & & $\pm 0.048$ & & $\pm 0.082$ \\
 & $\psi$ & $0.005$ & $\pm^{0}_{0.001}$ & $0.464$ & $\pm^{0.011}_{0.013}$ & $1.912$ & $\pm^{0.059}_{0.078}$ \\
 & & & $\pm^{0}_{0}$ & & $\pm^{0.001}_{0.003}$ & & $\pm^{0.009}_{0.005}$ \\ \cline{2-8}

 & & & $\pm 0.032$ & & $\pm 0.05$ & & $\pm 0.072$ \\
 & $\eta$ & $-0.034$ & $\pm^{0.005}_{0.004}$ & $0.372$ & $\pm^{0.009}_{0.011}$ & $1.809$ & $\pm^{0.055}_{0.071}$ \\
 & & & $\pm^{0}_{0}$ & & $\pm^{0.001}_{0.002}$ & & $\pm^{0.008}_{0.006}$ \\ \cline{2-8}

 & & & $\pm 0.022$ & & $\pm 0.057$ & & $\pm 0.056$ \\
 & $LR$ & $0.208$ & $\pm^{0.006}_{0.008}$ & $0.135$ & $\pm^{0.004}_{0.004}$ & $2.053$ & $\pm^{0.074}_{0.094}$ \\
 & & & $\pm^{0}_{0.001}$ & & $\pm^{0.001}_{0.001}$ & & $\pm^{0.008}_{0.007}$ \\ \hline \hline

 \multirow{12}{*}{3 TeV} & & & $\pm 0.290$ & & $\pm 0.369$ & & $\pm 2.81$ \\
 & $\chi$ & $-0.221$ & $\pm^{0.071}_{0.063}$ & $0.278$ & $\pm^{0.020}_{0.024}$ & $8.95$ & $\pm^{1.67}_{1.81}$ \\
 & & & $\pm^{0.001}_{0}$ & & $\pm^{0.001}_{0.001}$ & & $\pm^{0.04}_{0.02}$ \\ \cline{2-8}

 & & & $\pm 0.371$ & & $\pm 0.322$ & & $\pm 2.92$ \\
 & $\psi$ & $0.007$ & $\pm^{0.001}_{0.001}$ & $0.551$ & $\pm^{0.018}_{0.026}$ & $7.38$ & $\pm^{0.052}_{0.068}$ \\
 & & & $\pm^{0}_{0}$ & & $\pm^{0.01}_{0.01}$ & & $\pm^{0.08}_{0.07}$ \\ \cline{2-8}

 & & & $\pm 0.336$ & & $\pm 0.342$ & & $\pm 2.54$ \\
 & $\eta$ & $-0.027$ & $\pm^{0.015}_{0.018}$ & $0.465$ & $\pm^{0.015}_{0.023}$ & $7.08$ & $\pm^{0.47}_{0.59}$ \\
 & & & $\pm^{0}_{0}$ & & $\pm^{0.001}_{0.001}$ & & $\pm^{0.008}_{0.007}$ \\ \cline{2-8}

 & & & $\pm 0.239$ & & $\pm 0.406$ & & $\pm 2.04$ \\
 & $LR$ & $0.231$ & $\pm^{0.013}_{0.020}$ & $0.192$ & $\pm^{0.007}_{0.010}$ & $7.81$ & $\pm^{0.71}_{0.90}$ \\
 & & & $\pm^{0}_{0}$ & & $\pm^{0.001}_{0.001}$ & & $\pm^{0.008}_{0.007}$ \\ \hline
\end{tabular}
\end{center}
\caption{Forward-backward asymmetries and central to forward ratio for our example $Z'$ models.  From top to bottom, the errors are statistical ($100 \, {\rm fb}^{-1}$), PDF, and scale.}
\label{asym}
\end{table}

\begin{table}[ht]
\begin{center}
\begin{tabular}{| c | c || c l | c l | c l |}
\hline
 $M_{Z'}$ & Model & \multicolumn{2}{|c|}{$A_{FB}^{0.4}$} & \multicolumn{2}{|c|}{$A_{FB}^{off-peak,0.4}$} & \multicolumn{2}{|c|}{$R_{0.4}$} \\
\hline \hline
 \multirow{12}{*}{3 TeV} & & & $\pm 0.137$ & & $\pm 0.224$ & & $\pm 0.234$ \\
 & $\chi$ & $-0.225$ & $\pm^{0.042}_{0.035}$ & $0.200$ & $\pm^{0.016}_{0.016}$ & $1.231$ & $\pm^{0.122}_{0.133}$ \\
 & & & $\pm^{0.001}_{0}$ & & $\pm^{0.001}_{0.001}$ & & $\pm^{0.005}_{0.004}$ \\ \cline{2-8}

 & & & $\pm 0.186$ & & $\pm 0.211$ & & $\pm 0.282$ \\
 & $\psi$ & $0.005$ & $\pm^{0}_{0.001}$ & $0.461$ & $\pm^{0.025}_{0.027}$ & $1.096$ & $\pm^{0.054}_{0.062}$ \\
 & & & $\pm^{0}_{0}$ & & $\pm^{0}_{0.001}$ & & $\pm^{0.006}_{0.005}$ \\ \cline{2-8}

 & & & $\pm 0.170$ & & $\pm 0.218$ & & $\pm 0.252$ \\
 & $\eta$ & $-0.034$ & $\pm^{0.010}_{0.010}$ & $0.386$ & $\pm^{0.019}_{0.023}$ & $1.067$ & $\pm^{0.051}_{0.059}$ \\
 & & & $\pm^{0}_{0}$ & & $\pm^{0}_{0.015}$ & & $\pm^{0.007}_{0.005}$ \\ \cline{2-8}

 & & & $\pm 0.118$ & & $\pm 0.249$ & & $\pm 0.188$ \\
 & $LR$ & $0.201$ & $\pm^{0.013}_{0.019}$ & $0.152$ & $\pm^{0.007}_{0.009}$ & $1.138$ & $\pm^{0.068}_{0.076}$ \\
 & & & $\pm^{0}_{0}$ & & $\pm^{0}_{0}$ & & $\pm^{0.006}_{0.004}$ \\ \hline 
\end{tabular}
\end{center}
\label{results_0.4}
\caption{Forward-backward asymmetries and central to forward ratio for $y_1 = 0.4$ and $M_{Z'} = 3$ TeV.}
\end{table}

\section{Measuring charges \label{cmeas}}

Ideally, one would like to be able to measure the couplings directly, rather than refer to a particular model.  We study the extraction of four coupling 
combinations that can be determined from the on-peak observables considered above.  Two of the combinations are the parity symmetric 
combinations $c_u$, $c_d$ introduced in~\cite{Carena:2004xs}, which we introduce explicitly later.  We extend that parametrization to study the extraction of 
parity violating coupling combinations.

First, we note that the differential cross section can be expressed as
\be
\frac{d^{2}\sigma}{dy d\cos\theta} = \sum_{q=u,d} [a_1^{q'} (q_R^2 + q_L^2)(e_R^2 + e_L^2) + a_2^{q'} (q_R^2 - q_L^2)(e_R^2 - e_L^2)] .
\label{sigold}
\ee
We have assumed that the couplings are generation-independent, thus the contributions from different quarks of the same type can be combined.  The coefficients 
$a_1^{q'}$ and $a_2^{q'}$ represent what is left after the couplings are extracted from the differential cross section.  They are composed of the PDFs 
and matrix elements integrated over phase space subject to the cuts discussed above.  We have checked that the $\gamma$ and $Z$ interference and squared terms 
are negligible, and we have dropped them in this parameterization.

Let us define, as in \cite{Carena:2004xs}, the parity symmetric coupling combinations
\be
c_q = \frac{M_{Z'}}{24\pi\Gamma}(q_R^2 + q_L^2)(e_R^2 + e_L^2) = (q_R^2 + q_L^2) Br(Z' \ra e^+ e^-) .
\ee
We further define the parity-violating combinations
\be
e_q = \frac{M_{Z'}}{24\pi\Gamma}(q_R^2 - q_L^2)(e_R^2 - e_L^2),
\ee
which can be accessed via measurements of the differential cross section.  Our equation for the differential cross section now reads
\be
\frac{d^{2}\sigma}{dy d\cos\theta} = \sum_{q=u,d} \frac{24 \pi \Gamma}{M_{Z'}} [a_1^{q'} c_q + a_2^{q'} e_q] .
\label{sig}
\ee
We absorb the overall factor into the coefficients, $a_{1,2}^{q} = \frac{24 \pi \Gamma}{M_{Z'}} a_{1,2}^{q'}$, so that
\be
\frac{d^{2}\sigma}{dy d\cos\theta} = \sum_{q=u,d} [a_1^{q} c_q + a_2^{q} e_q] .
\label{sig2}
\ee
In the narrow width approximation, the $a_{1,2}^{q'}$ scale as $a_{1,2}^{q'} \sim 1/\Gamma$.  Switching to $a_{1,2}^{q}$ removes almost all width 
dependence from these factors; we have checked that they vary by less than $0.5\%$ over the range of widths considered here.  The only dependence of the 
$a_{1,2}^{q}$ on the $Z'$ being considered is through $M_{Z'}$.  These coefficients therefore need be computed only once for a given $M_{Z'}$ and set of cuts.  
We present them below for $M_{Z'}=1.5$ TeV.

We propose to use Eq. \ref{sig2} to determine the four quantities $c_q$ and $e_q$.  By integrating Eq. \ref{sig2} over four different regions in $y$ and $\theta$, one obtains four equations for the four unknowns.  The left-hand-side is determined by experiment and the integrated coefficients $a_1$, $a_2$ are determined theoretically.  Thus one can solve for the unknown $c_q$ and $e_q$.  While in principle any four independent regions could be used, we would like to minimize the errors by isolating each of the four unknowns as much as possible in our four equations.  To do this, we use the four observables $F_< = \int_{-y_1}^{y_1} dy F(y), B_< = \int_{-y_1}^{y_1} dy B(y), F_> = (\int_{y_1}^{y_{max}} + \int_{-y_{max}}^{y_1}) dy F(y),$ and $B_> = (\int_{y_1}^{y_{max}} + \int_{-y_{max}}^{y_1}) dy B(y)$.  Comparing $F$ and $B$ helps to separate $c_q$ from $e_q$; one expects $c_q$ to contribute to $F+B$ and $e_q$ to $F-B$.  Separating different $Z'$ rapidities helps to isolate up-type from down-type couplings due to their different PDFs.  These quantities 
are related to the forward-backward asymmetry, rapidity ratio, and cross section via
\be
F_< = \frac{\sigma}{2}(1 + A_{FB}^{0} - \frac{1+A_{FB}^{y_1}}{R_{y_1}+1}), \\
B_< = \frac{\sigma}{2}(1 - A_{FB}^{0} - \frac{1-A_{FB}^{y_1}}{R_{y_1}+1}), \\
F_> = \frac{\sigma}{2}(\frac{1+A_{FB}^{y_1}}{R_{y_1}+1}), \\
B_> = \frac{\sigma}{2}(\frac{1-A_{FB}^{y_1}}{R_{y_1}+1}) .
\ee

After integrating Eq. \ref{sig2} over these four regions, we end up with the system of linear equations
\be
\vec{m} = \mathbf{M} \vec{c}
\ee
where
\be
\vec{m} = \left( \begin{array}{c} F_< \\ B_< \\ F_> \\ B_> \\ \end{array} \right),
\vec{c} = \left( \begin{array}{c} c^u \\ c^d \\ e^u \\ e^d \\ \end{array} \right),
\ee
and $\mathbf{M}$ is a matrix composed of the coefficients $a_1^q$ and $a_2^q$ integrated over the appropriate ranges of $y$ and $\theta$.  Explicitly, 
$\mathbf{M}$ takes the form
\be
\mathbf{M} = \left( \begin{array}{c c c c}
	\int_{F_<} a_1^u & \int_{F_<} a_1^d & \int_{F_<} a_2^u & \int_{F_<} a_2^d \\
	\int_{B_<} a_1^u & \int_{B_<} a_1^d & \int_{B_<} a_2^u & \int_{B_<} a_2^d \\
	\int_{F_>} a_1^u & \int_{F_>} a_1^d & \int_{F_>} a_2^u & \int_{F_>} a_2^d \\
	\int_{B_>} a_1^u & \int_{B_>} a_1^d & \int_{B_>} a_2^u & \int_{B_>} a_2^d \\
	\end{array} \right)
\ee
The entries are determined by running our code for certain basis models with charges such that one of the $c_q$ or $e_q$ are equal to 1, 
and the others are zero.  The photon and Z contributions have been checked to be negligible and are turned off in this process.  
For example, with $y_1 = 0.8$, using the central PDF, we get 
\be
\mathbf{M} = \left( \begin{array}{c c c c}
	5638 & 4175 & 1747 & 828 \\
	5638 & 4175 & -1746 & -827 \\
	3610 & 1519 & 2101 & 784 \\
	3610 & 1519 & -2101 & -784 \\
	\end{array} \right) {\rm fb}
\ee
for $M_{Z'} = 1.5$ TeV.  Solving for the couplings is now straightforward: $\vec{c} = \mathbf{M}^{-1} \vec{m}$.  As mentioned we have found that these entries vary by less 
than 0.5\% over the range of widths for the examined models.

Below we illustrate this extraction procedure applied to our four example models.  We simulate experimental results 
for $F_<$, $B_<$, $F_>$, and $B_>$ using the full NLO cross section {\it including} the $\gamma$ and $Z$ contributions.  We assume an integrated 
luminosity of $100 \, {\rm fb}^{-1}$.  We then subject these results to our extraction procedure described above.  Doing so, we 
obtain ``measurements'' of $c_q$ and $e_q$.  These measurements are compared to the theoretical input parameters used in generating the data.  Our results 
are shown below in Table~\ref{couplings}.  For $M_{Z'}=1.5$ TeV we use $y_1=0.8$; for $3$ TeV we use $y_1=0.4$.  We have checked that scale errors are negligible in this analysis, as discussed previously, and they have been omitted.

\begin{table}[h]
\begin{center}
\begin{tabular}{| c | c || c | c | c | c |}
\hline
 $M_{Z'}$ & Mdl & $c_u \times 10^3$ & $c_d \times 10^3$ & $e_u \times 10^3$ & $e_d \times 10^3$ \\ \hline \hline
 & & $0.66$ & $3.30$ & $0$ & $-2.11$ \\
 & \raisebox{1.5ex}{$\chi$} & $0.68 \pm 0.14 \pm^{0.25}_{0.32}$ & $3.32 \pm 0.24 \pm^{0.58}_{0.47}$ & $0.01 \pm 0.76 \pm^{0.72}_{0.35}$ & $-2.09 \pm 1.92 \pm^{0.82}_{1.70}$ \\ \cline{2-6}
 & & $0.81$ & $0.81$ & $0$ & $0$ \\
 1.5 & \raisebox{1.5ex}{$\psi$} & $0.81 \pm 0.10 \pm^{0.06}_{0.08}$ & $0.81 \pm 0.17 \pm^{0.19}_{0.15}$ & $0.01 \pm 0.53 \pm^{0.01}_{0}$ & $0.00 \pm 1.32 \pm^{0.01}_{0.01}$ \\ \cline{2-6}
 TeV & & $1.08$ & $0.67$ & $0$ & $-0.24$ \\
 & \raisebox{1.5ex}{$\eta$} & $1.09 \pm 0.11 \pm^{0.07}_{0.09}$ & $0.68 \pm 0.18 \pm^{0.20}_{0.16}$ & $0.01 \pm 0.57 \pm^{0.08}_{0.04}$ & $-0.23 \pm 1.41 \pm^{0.09}_{0.19}$ \\ \cline{2-6}
 & & $1.59$ & $2.69$ & $0.60$ & $1.03$ \\
 & \raisebox{1.5ex}{$LR$} & $1.61 \pm 0.16 \pm^{0.19}_{0.26}$ & $2.71 \pm 0.26 \pm^{0.53}_{0.41}$ & $0.57 \pm 0.84 \pm^{0.16}_{0.33}$ & $1.12 \pm 2.10 \pm^{0.83}_{0.42}$ \\ \hline \hline
 & & $0.67$ & $3.36$ & $0$ & $-2.15$ \\
 & \raisebox{1.5ex}{$\chi$} & $0.69 \pm 1.4 \pm^{0.59}_{1.15}$ & $3.36 \pm 3.02 \pm^{2.22}_{1.23}$ & $0.00 \pm 3.26 \pm^{0.62}_{0.35}$ & $-2.12 \pm 7.81 \pm^{0.57}_{1.12}$ \\ \cline{2-6}
 & & $0.82$ & $0.82$ & $0$ & $0$ \\
 3 & \raisebox{1.5ex}{$\psi$} & $0.82 \pm 1.01 \pm^{0.18}_{0.28}$ & $0.82 \pm 2.22 \pm^{0.58}_{0.42}$ & $0.01 \pm 2.36 \pm^{0}_{0}$ & $-0.00 \pm 5.62 \pm^{0}_{0}$ \\ \cline{2-6}
 TeV & & $1.10$ & $0.69$ & $0$ & $-0.25$ \\
 & \raisebox{1.5ex}{$\eta$} & $1.09 \pm 1.10 \pm^{0.22}_{0.28}$ & $0.69 \pm 2.41 \pm^{0.61}_{0.51}$ & $0.00 \pm 2.55 \pm^{0.07}_{0.04}$ & $-0.25 \pm 6.07 \pm^{0.06}_{0.12}$ \\ \cline{2-6}
 & & $1.62$ & $2.74$ & $0.61$ & $1.05$ \\
 & \raisebox{1.5ex}{$LR$} & $1.63 \pm 1.6 \pm^{0.50}_{0.89}$ & $2.74 \pm 3.45 \pm^{1.79}_{1.09}$ & $0.57 \pm 3.68 \pm^{0.14}_{0.18}$ & $1.09 \pm 8.78 \pm^{0.53}_{0.29}$ \\ \hline
\end{tabular} 
\end{center}
\caption{Result of extracting $Z'$ couplings at the LHC. Theoretical values for the couplings are listed first.  The results of our 
extraction procedure are shown next.  Errors are statistical and PDF, respectively.}
\label{couplings}
\end{table}
We see that there is good agreement between the couplings extracted using this method and the theoretical input values.  One caveat must be discussed.  
Clearly, our simulation of the experimental results is simplistic.  However, it allows us to study two important issues.  First, it shows us that 
this technique is consistent, and particularly that the neglect of the $\gamma$ and $Z$ terms in Eq.~\ref{sig2} is justified.  It also allows us 
to study how accurately the $c_q$ and $e_q$ can be determined at the LHC given the expected theoretical and statistical errors.  For $100 \, {\rm fb}^{-1}$, individual errors on $c_q$ and $e_q$ can be rather large, especially in the $3$ TeV case.  However, we will see that when the couplings are taken together, models are well-discriminated, especially at the SLHC.

\begin{figure}[htbp]
\centerline{
\includegraphics[height=12.0cm,width=9.0cm,angle=90]{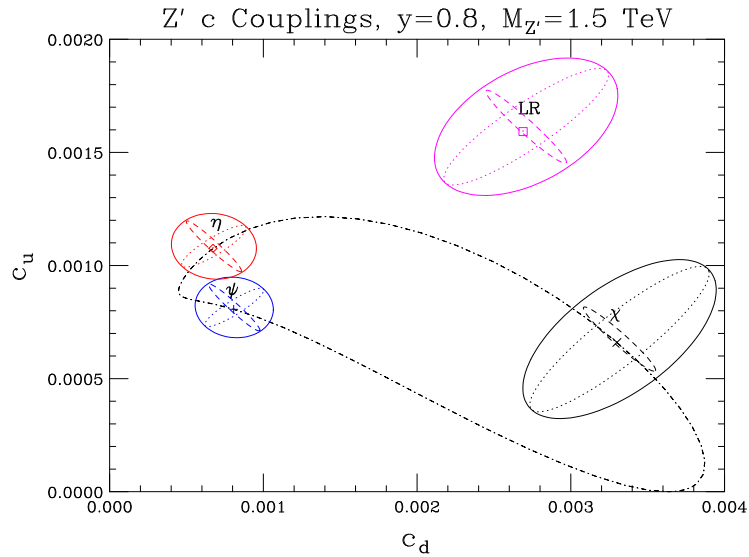}
}
\vspace{-0.9cm}
\caption{Simulated measurements of the $c_{u,d}$ couplings at the LHC for our test models.  The dashed ellipses are the statistical errors expected for
$M_{Z'} = 1.5$ TeV and $100 \, {\rm fb}^{-1}$ of data, the dotted ellipses are the current estimated PDF errors, and the solid ellipses denote the 
combined errors.  The $E_6$ family of models lie on the dot-dashed contour.}
\label{cplot}
\end{figure}

\begin{figure}[htbp]
\centerline{
\includegraphics[height=12.0cm,width=9.0cm,angle=90]{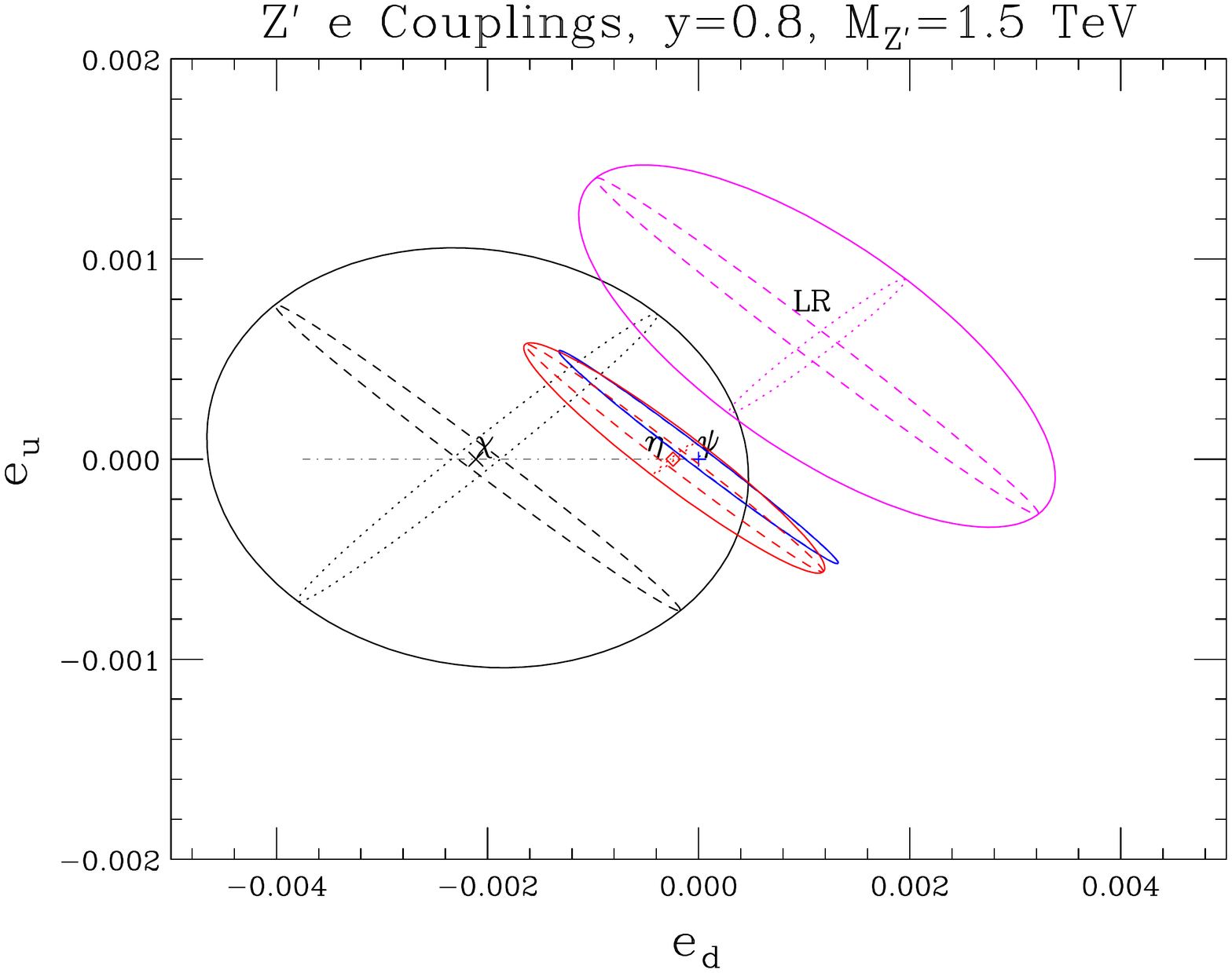}
}
\vspace{-0.9cm}
\caption{Simulated measurements of the $e_{u,d}$ couplings at the LHC for our test models.  The parameters and ellipses are as discussed in the previous
plot caption; the $E_6$ contour is a line segment.}
\label{eplot}
\end{figure}

\begin{figure}[htbp]
\centerline{
\includegraphics[height=12.0cm,width=9.0cm,angle=90]{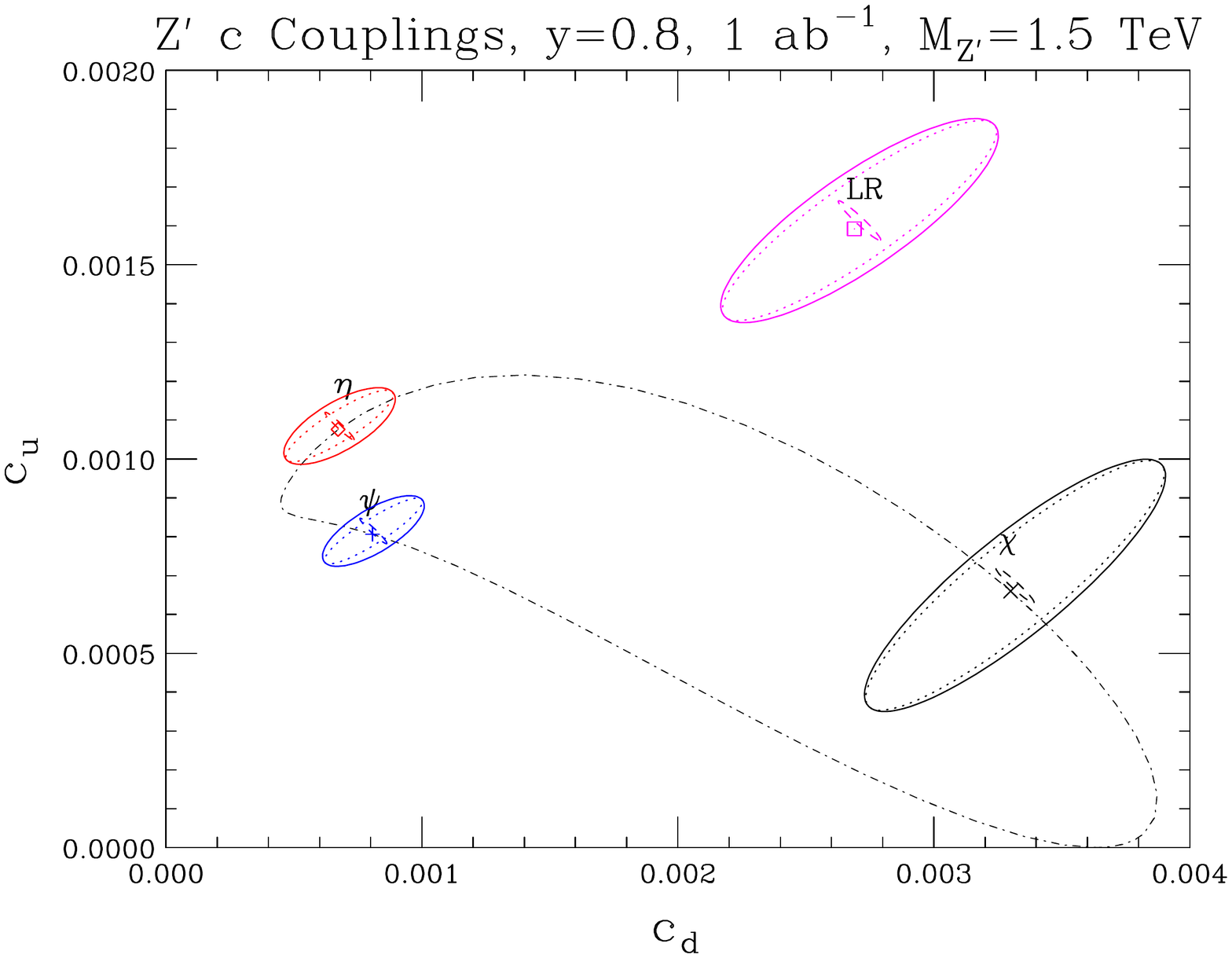}
}
\vspace{-0.9cm}
\caption{Simulated measurements of the $c_{u,d}$ couplings at the SLHC for our test models.  The dashed ellipses are the statistical errors expected for
$M_{Z'} = 1.5$ TeV and $1 \, {\rm ab}^{-1}$ of data, the dotted ellipses are the current estimated PDF errors, and the solid ellipses denote the 
combined errors.    The $E_6$ family of models lie on the dot-dashed contour.}
\label{cplot2}
\end{figure}

\begin{figure}[htbp]
\centerline{
\includegraphics[height=12.0cm,width=9.0cm,angle=90]{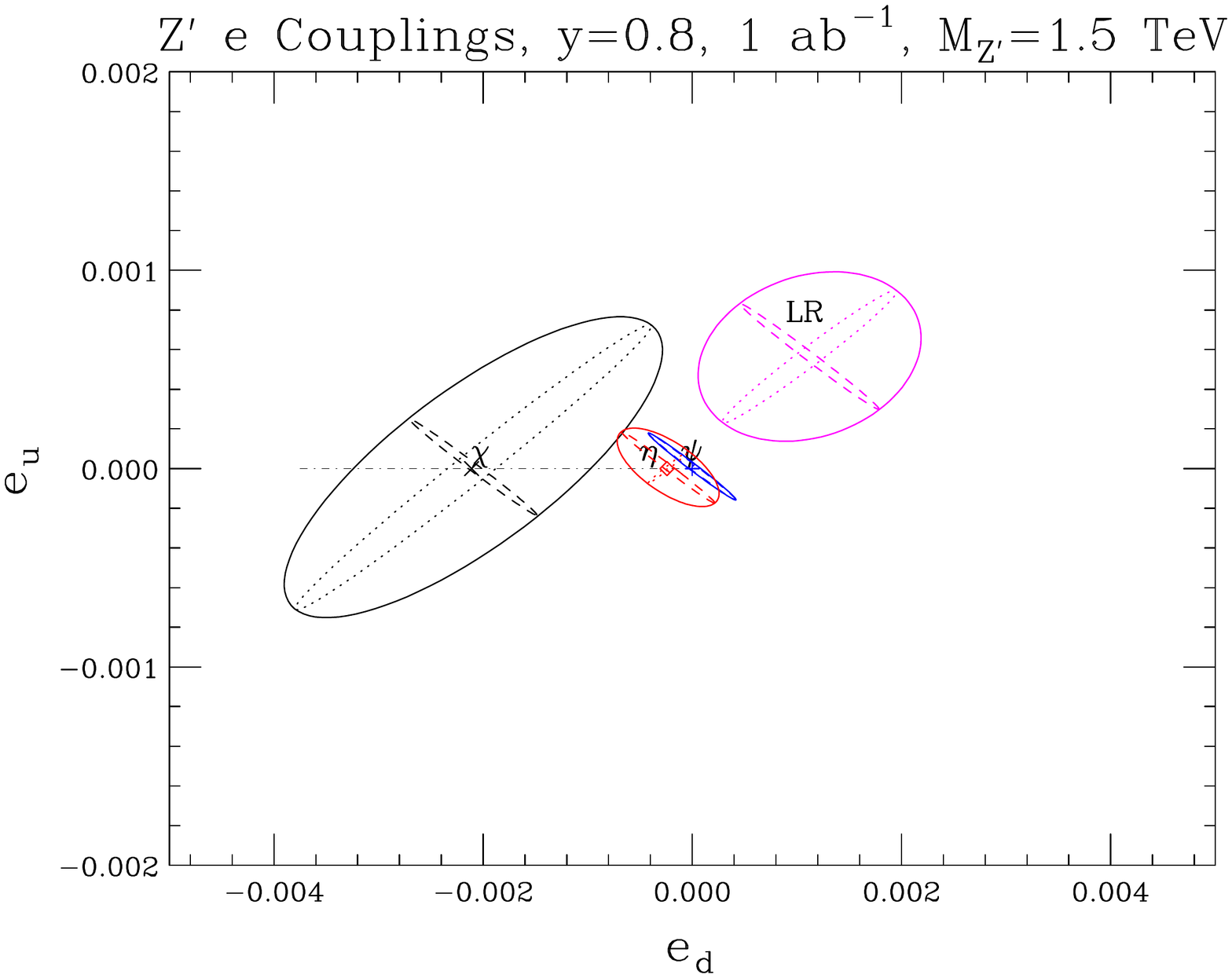}
}
\vspace{-0.9cm}
\caption{Simulated measurements of the $e_{u,d}$ couplings at the SLHC for our test models.  The parameters and ellipses are as discussed in the previous
plot caption; the $E_6$ contour is a line segment.}
\label{eplot2}
\end{figure}

\begin{figure}[htbp]
\centerline{
\includegraphics[height=12.0cm,width=9.0cm,angle=90]{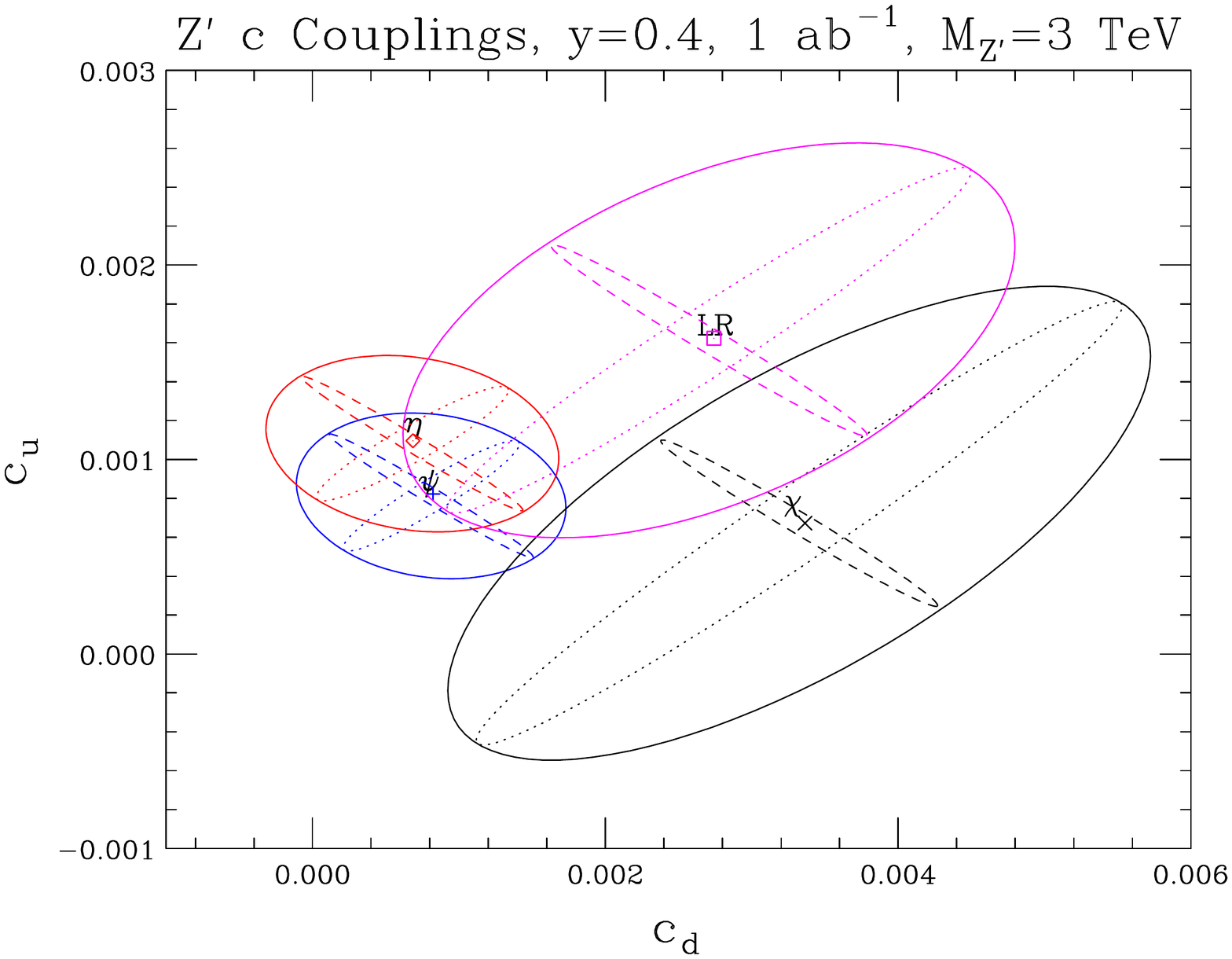}
}
\vspace{-0.9cm}
\caption{Simulated measurements of the $c_{u,d}$ couplings at the LHC for our test models.  The dashed ellipses are the statistical errors expected for
$M_{Z'} = 3$ TeV and $1 \, {\rm ab}^{-1}$ of data, the dotted ellipses are the current estimated PDF errors, and the solid ellipses denote the 
combined errors.}
\label{cplot3}
\end{figure}

\begin{figure}[htbp]
\centerline{
\includegraphics[height=12.0cm,width=9.0cm,angle=90]{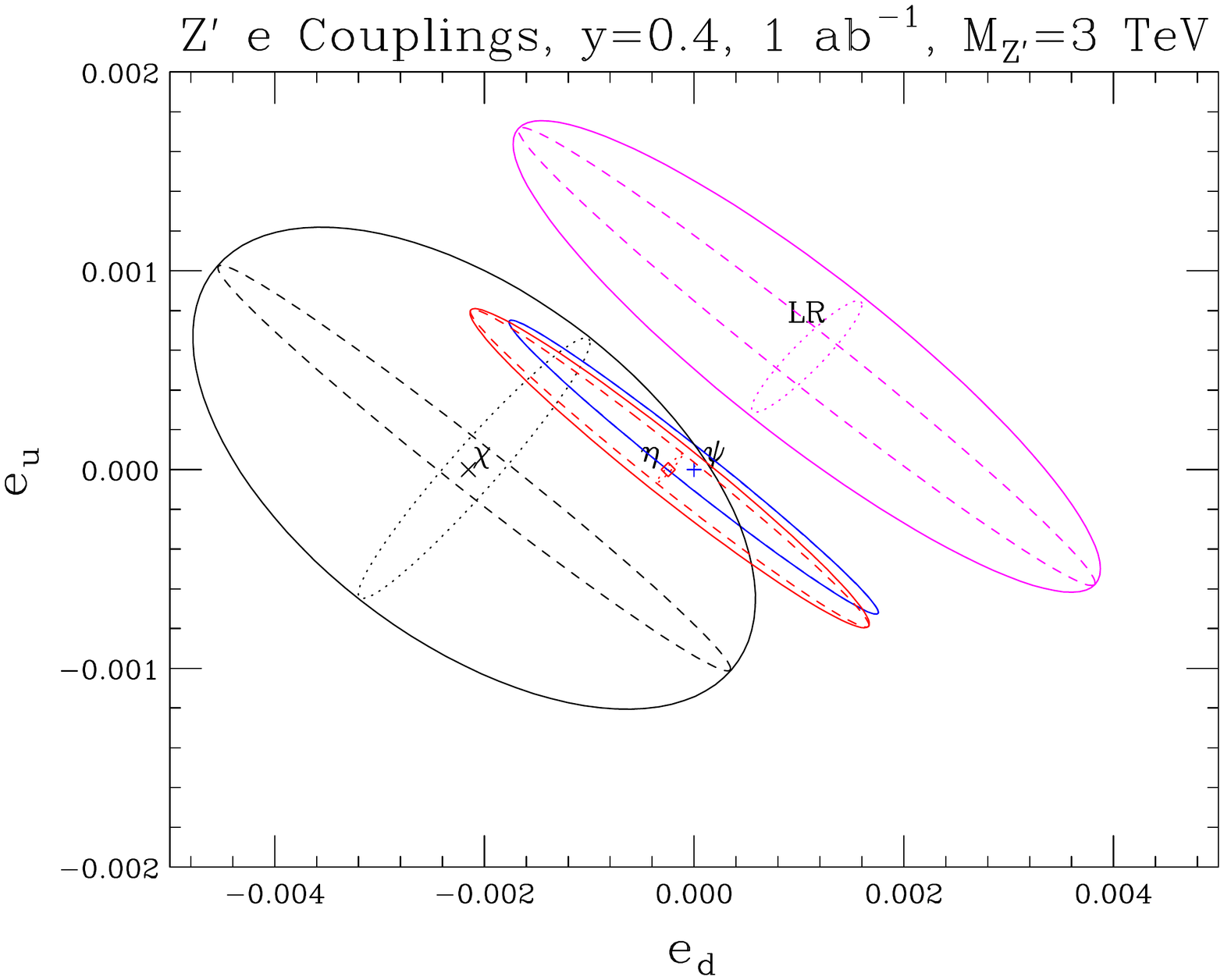}
}
\vspace{-0.9cm}
\caption{Simulated measurements of the $e_{u,d}$ couplings at the SLHC for our test models at $M_{Z'}=3$ TeV.  The parameters and ellipses are as discussed in the previous
plot caption.}
\label{eplot3}
\end{figure}

We now illustrate these results graphically for $M_{Z'} = 1.5$ TeV, first for the LHC assuming $100 \, {\rm fb}^{-1}$ and then for the SLHC assuming 
$1 \, {\rm ab}^{-1}$.  The LHC results for $c_{u,d}$ and $e_{u,d}$ are projected into a $c_{u,d}$-plane and $e_{u,d}$-plane, and shown in Figs.~\ref{cplot} and~\ref{eplot}, respectively; SLHC results are in Figs.~\ref{cplot2} and~\ref{eplot2}.  We also present results for $M_{Z'} = 3$ TeV for the SLHC in Figs.~\ref{cplot3} and~\ref{eplot3}; the statistical errors at the LHC are simply too large for a meaningful coupling extraction without more data.  In general, the errors are ellipsoids in the four-dimensional space of $(c_u,c_d,e_u,e_d)$.  The statistical errors are very nearly diagonal in $c_u$/$c_d$ vs. $e_u$/$e_d$ due to our choice of measurements.  However, the PDF and therefore the combined errors are not.  For simplicity of presentation we plot the {\it projections} of the 4-D PDF and combined errors into the $c_{u,d}$ and $e_{u,d}$ planes.  We also conservatively take the larger of the +/- PDF deviations for each ellipse.  We have plotted a contour for the family of $E_6$ models on the $M_Z' = 1.5$ TeV plots for reference.  We note the following trends in these plots.
\begin{itemize}

\item The statistical error ellipses are rather narrow, with minimal extent in the $c_u+c_d$ and $e_u+e_d$ directions.  This occurs because statistical errors 
are reduced when we add contributions form up and down quarks, rather than attempt to distinguish between them.

\item The PDF error projections are nearly orthogonal to the statistical errors, and are minimized in the $c_u-c_d$ and $e_u-e_d$ directions.  We see that models are easily distinguishable with only statistical errors due to the narrow ellipses, but when including PDF errors, a large volume of coupling space is occupied.  However, for a $1.5$ TeV $Z'$, the $c$ couplings can still distinguish models.  One can still determine these coupling combinations within a reasonable window.

\item The PDF errors scale with the couplings.  For models with zero values, such as the $\psi$ model for $e_u$ and $e_d$, the errors are mostly statistical.

\item The $e$ couplings are harder to measure.  Much of this stems from the fact that the difference $F-B$ is used to extract them, which has higher statistical error than the corresponding $F+B$ for the $c$ couplings.

\item {\it All} $E_6$ models have $e_u = 0$ and $e_d \le 0$.  A substantial departure from this could rule out this family, though due to large statistical errors the SLHC might be needed.

\item The error ellipses are quite narrow in Figs.~\ref{cplot2} and ~\ref{eplot2}, for a $1.5$ TeV $Z'$ at the SLHC, allowing reasonable determination of the couplings.  The errors will improve further should PDF errors improve.

\item Charge extraction for $M_{Z'} = 3$ TeV is difficult, even at the SLHC with present PDF error estimates.

\end{itemize}




\subsection{Distinguishing Models}

We now test whether the above observables are sufficient to distinguish our example models at the LHC.  We compare models pairwise by assuming that one is correct and finding the chi-squared of the other model as a test, using the errors of the first model.  Statistical and PDF errors are combined in quadrature.  While the PDF errors roughly represent a 90\% confidence level, the precise meaning isn't clear~\cite{Owens}.  We conservatively take them to be $1 \sigma$ in the combination and again take the larger of the +/- deviations, as in our plots.  Scale errors are unimportant in this analysis and have been dropped.

We use the charges $c_q$ and $e_q$, as well as $A_{FB}^{off-peak, 0.4}$ to form the $\chi^{2}$ for $1.5$ TeV.  Rather than use the charges $c_q$ and $e_q$ directly, we form the $\chi^2$ by diagonalizing the errors in these variables, as in our plots.  This allows us to best exploit the model separation in directions where the errors are minimized.  $A_{FB}$ numbers tend to contribute the most to $\chi^{2}$ since their errors are relatively small, and this is the only off-peak data we have included here.  If one uses only $A_{FB}$, choosing a lower rapidity gives more data, reducing the error.  However, for too low a rapidity, quark direction misidentification tends to wash out the actual values.  We have found that $y_1=0.4$ is a good choice when not including other data.

\begin{table}[h]
\begin{center}
\begin{tabular}{| c | c || c | ccc | ccc | ccc | ccc |}
\hline
 & & & \multicolumn{3}{|c|}{$\chi$} & \multicolumn{3}{|c|}{$\psi$} & \multicolumn{3}{|c|}{$\eta$} 
& \multicolumn{3}{|c|}{$LR$} \\ 
 $M_{Z'}$ & Mdl & $y_1$ & $\chi^2_{c,e}$ & $\chi^2_{tot}$ & $\sigma$ & $\chi^2_{c,e}$ & $\chi^2_{tot}$ & $\sigma$ & $\chi^2_{c,e}$ & $\chi^2_{tot}$ & $\sigma$ & $\chi^2_{c,e}$ & $\chi^2_{tot}$ & $\sigma$ \\ \hline \hline
 & & 0.6 & & & & 252 & 300 & $17$ & 125 & 149 & $11$ & 35 & 35 & $4.8$ \\ 
 & $\chi$ & 0.8 & & & & 223 & 272 & $16$ & 125 & 150 & $11$ & 42 & 43 & $5.5$ \\
 & & 1.0 & & & & 207 & 256 & $15$ & 119 & 142 & $11$ & 45 & 45 & $5.7$ \\ \cline{2-15}
 & & 0.6 & 47 & 98 & $9.1$ & & & & 4.3 & 9.0 & 1.6 & 15 & 74 & $7.7$ \\
 & $\psi$ & 0.8 & 51 & 102 & $9.3$ & & & & 4.2 & 8.8 & 1.6 & 15 & 73 & $7.6$ \\
 1.5 & & 1.0 & 56 & 107 & $9.5$ & & & & 3.7 & 8.3 & 1.5 & 15 & 73 & $7.6$ \\ \cline{2-15}
 TeV & & 0.6 & 58 & 82 & $8.2$ & 7.6 & 12 & 2.1 & & & & 15 & 43 & $5.5$ \\
 & $\eta$ & 0.8 & 61 & 85 & $8.4$ & 6.8 & 11 & 2.0 & & & & 15 & 43 & $5.5$ \\ 
 & & 1.0 & 64 & 89 & $8.5$ & 6.9 & 11 & 2.0 & & & & 15 & 43 & $5.5$ \\ \cline{2-15}
 & & 0.6 & 15 & 15 & 2.6 & 159 & 201 & $14$ & 71 & 92 & $8.8$ & & &\\
 & $LR$ & 0.8 & 17 & 17 & $2.8$ & 187 & 230 & $15$ & 75 & 96 & $9.0$ & & &\\ 
 & & 1.0 & 16 & 17 & 2.8 & 174 & 217 & $14$ & 68 & 89 & $8.6$ & & &\\ \hline \hline
 & $\chi$ & 0.4 & & & & 6.7 & 8.1 & 1.4 & 5.7 & 6.3 & 1.1 & 11 & 11 & 2.0 \\ \cline{2-15}
 3 & $\psi$ & 0.4 & 3.7 & 5.2 & 0.9 & & & & 0.3 & 0.4 & 0.1 & 3.5 & 5.7 & 1.0 \\
 \cline{2-15}
 TeV & $\eta$ & 0.4 & 2.7 & 3.5 & 0.5 & 0.2 & 0.3 & 0.0 & & & & 3.7 & 4.8 & 0.8 \\ 
 \cline{2-15}
 & $LR$ & 0.4 & 3.9 & 4.0 & 0.6 & 11 & 12 & 2.1 & 8.1 & 9.0 & 1.6 & & & \\ 
 \hline
\end{tabular}
\end{center}
\caption{Pairwise $\chi^{2}$ values for our model comparison, for $100\, {\rm fb}^{-1}$.  The separate $\chi^2$ contributions from the on-peak $c_q,e_q$ couplings have been shown, as have the total $\chi^2$ values including $A_{FB}^{off-peak, 0.4}$, and the corresponding confidence that the models are distinct in standard deviations.  The model in each row is assumed to be the correct, measured model, and is tested against the hypothesis in each column.  We have performed this test for several choices of on-peak $y_1$; for $M_{Z'}=1.5$ TeV, $y_1=0.8$ appears to be the optimal choice.  For the off-peak asymmetry, only $y_1=0.4$ has been used.  For $M_{Z'}=3$ TeV only $y_1=0.4$ has been used.  Note that since the statistical errors come from the row models, and PDF from the columns, this table is not symmetric.}
\label{chisq} 
\end{table}

We see from Table \ref{chisq} that for $M_{Z'} = 1.5$ TeV, these observables should distinguish the considered models quite reliably.  As values of $1-2\,\sigma$ 
correspond to confidence levels of 68\% and 95\% respectively, we see that most are distinguishable with 99\% C.L. or greater.  While the $\psi$ and $\eta$ models have similar values for all observables, the errors are small enough that together one can separate them with a confidence level of nearly 90\%.  The other models are easily distinguishable with $100 \,{\rm fb}^{-1}$ of data.  In the full $c/e$ analysis, $y_1=0.8$ appears to be the most discriminating choice.  For the heavy $3$ TeV $Z'$, there are very few high rapidity events; we have therefore restricted our analysis to $y_1=0.4$.  However, there is still some distinguishability at $100 \, {\rm fb}^{-1}$ with this choice, despite the large errors seen in the plots and in Table~\ref{couplings}.  Several models can be separated at $1-2 \sigma$ (68-95\% CL).  We see below in Table~\ref{chisq2} that the situation improves signficantly with the SLHC, with all models but the $\psi/\eta$ pair well-separated.

\begin{table}[h]
\begin{center}
\begin{tabular}{| c | c || ccc | ccc | ccc | ccc |}
\hline
 & & \multicolumn{3}{|c|}{$\chi$} & \multicolumn{3}{|c|}{$\psi$} & \multicolumn{3}{|c|}{$\eta$} 
& \multicolumn{3}{|c|}{$LR$} \\ 
 $M_{Z'}$ & Mdl & $\chi^2_{c,e}$ & $\chi^2_{tot}$ & $\sigma$ & $\chi^2_{c,e}$ & $\chi^2_{tot}$ & $\sigma$ & $\chi^2_{c,e}$ & $\chi^2_{tot}$ & $\sigma$ & $\chi^2_{c,e}$ & $\chi^2_{tot}$ & $\sigma$ \\ \hline \hline
 & $\chi$ & & & & 49 & 61 & $6.8$ & 37 & 43 & $5.5$ & 32 & 32 & $4.5$ \\ \cline{2-14}
 3 & $\psi$ & 15 & 29 & $4.3$ & & & & 1.1 & 2.3 & $0.2$ & 4.6 & 26 & $3.9$ \\
 \cline{2-14}
 TeV & $\eta$ & 15 & 22 & $3.4$ & 1.3 & 2.3 & $0.2$ & & & & 13 & 24 & $3.7$ \\ 
 \cline{2-14}
 & $LR$ & 14 & 14 & $2.4$ & 44 & 58 & $6.7$ & 30 & 38 & $5.1$ & & & \\ 
 \hline
\end{tabular} 
\end{center}
\caption{Pairwise $\chi^{2}$ for $1\,{\rm ab}^{-1}$, $y_1=0.4$, and $M_{Z'}=3$ TeV.  As before, the rows are tested against the hypothesis columns.}
\label{chisq2}
\end{table}

In principle, one could use the off-peak data in a full analysis of the charges by modifying Eq. \ref{sigold} and adding the neglected Z and photon contributions.  The expression for 
the differential cross section becomes
\be
\frac{d^{2}\sigma}{dy d\cos\theta} & = & \sum_{q=u,d} [a_1^q (q_R^2 + q_L^2)(e_R^2 + e_L^2) + a_2^q (q_R^2 - q_L^2)(e_R^2 - e_L^2) + b_1^q q_R e_R + b_2^q q_R e_L] \nonumber \\
& & \mbox{} + b_3 q_L e_R + b_4 q_L e_L + c .
\label{sigoff}
\ee
The known Z and photon charges have been folded into the coefficients $b$ and $c$.  There are five unknowns, $q_{L,R}$ and $e_{L,R}$, but only four independent combinations of these appear above.  This can be seen by noting that  increasing the quark charges by a factor of two and decreasing the lepton couplings by a factor of two leaves the differential cross section unchanged, indicating that a degeneracy exists.  Modifying our above procedure would require solving four quartic equations for the four independent unknowns.  However, the linear terms in $q \times e$ provide information on the signs of the charges, which is an enticing prospect; it is likely one can gain significant statistical precision, and determine the signs with confidence.  In addition, if one can determine the invisible width, this would also yield $Br(Z' \ra e^+ e^-)$, which could be used to separate $q \times e$ and solve for the individual charges $q_{L,R}$ and $e_{L,R}$, with signs if off-peak data is analyzed.  The ILC should also be able to probe the interference region for the masses considered \cite{Freitas:2003yp}, and may also be able to determine $e_{L,R}$ directly, breaking our $q \times e$ degeneracy.

\section{Conclusions \label{conc}}

In this paper we studied the measurement of $Z'$ couplings at the LHC.  We performed a fully differential NLO QCD calculation of the $Z'$ signal with all 
spin correlations and interference effects with the SM $\gamma$ and $Z$ included, as well as realistic LHC acceptance cuts.  
Using four example models arising from grand unified theories, we quantified the effect of statistical, PDF, and residual 
scale errors on important $Z'$ observables.  We found that residual scale errors are negligible for observables formed from cross section 
ratios, such as $A_{FB}$, indicating that corrections from NNLO QCD effects are unimportant in $Z'$ studies.  However, statistical and PDF errors are 
significant and have approximately equal effects with $100 \, {\rm fb}^{-1}$ at the LHC.

We introduced a set of $Z'$ coupling combinations that can be determined from on-peak measurements at the LHC.  The parity symmetric 
combinations $c_{u,d}$ which can be accessed by measurements of the inclusive cross section were introduced previously 
in~\cite{Carena:2004xs}.  We extended this parametrization to include the parity violating combinarions $e_{u,d}$ that can 
be probed once differential measurements are made.  The differential cross section factors into a sum over products of these couplings 
times transfer functions that depend on the model under consideration only through the $Z'$ mass; the slight dependence of these functions 
on the width was found to be less than 1\%.  These transfer functions only need to be evaluated once for a given $Z'$ mass and set of cuts, and then 
can then be used in simulations regardless of the underlying $Z'$ model.  They form a matrix which connects measurements in different 
kinematic regions to the underlying $Z'$ couplings.  To access these transfer function one needs only to run a $Z'$ simulation code for basis 
vectors in coupling space of the form $(c_u,c_d,e_u,e_d)=(1,0,0,0)$, etc.

We computed these transfer functions and used them to examine how well $Z'$ couplings can be determined at the LHC assuming $100 \, {\rm fb}^{-1}$, and at the 
SLHC assuming $1 \, {\rm ab}^{-1}$.  As illustrative examples we again used the four example models discussed previously.  Both statistical and PDF 
errors 
give equally important contributions to the uncertainty in coupling measurements at the LHC.  The statistical and PDF errors turn out to be maximal 
in orthogonal directions in both the $(c_u,c_d)$ and $(e_u,e_d)$ planes.  We found that the $c_{u,d}$ could be determined with reasonably good precision 
at the LHC, easily well enough to distinguish between the four example models considered.  However, the $e_{u,d}$ will be relatively poorly determined 
with $100 \, {\rm fb}^{-1}$ assuming current PDF errors; a more accurate measurement of these couplings sufficient to tell which of the 
four example models they came from required $1 \, {\rm ab}^{-1}$ in our analysis.

Our analysis can be extended in several ways.  Inclusion of off-peak observables can improve the 
precision of the coupling extractions; we saw in Section~\ref{conc} that the off-peak asymmetry 
increased the $\chi^2$ in the point-wise comparison between models.  However, off the $Z'$ peak 
interference terms between the $Z'$ and SM $\gamma$ and $Z$ give the dominant contributions.  Our 
coupling parametrization must be enlarged to include these effects.  Another direction in which to 
extend our analysis would be to break the degeneracy between lepton and quark couplings to the 
$Z'$.  This degeneracy is clearly visible in Eqs.~\ref{sigold} and~\ref{sigoff}; if we scale the quark couplings up 
by a factor of two; scaling the lepton couplings down by the same factor leads to an unchanged cross section.  
Measurements of rare $Z'$ decays can break this degeneracy~\cite{delAguila:1993ym}.  In principle the $Z'$ width could break this 
degeneracy if the invisible decay width was known.  The $Z'$ invisible width is also of interest in models where the 
$Z'$ acts as the messenger to a hidden sector, as it gives some insight into the matter content of this sector of the theory.

\vspace{0.2cm}
\noindent
{\bf Acknowledgments:} 
It is a pleasure to thank T. Rizzo for helpful comments on an initial version of this manuscript.  The authors are supported by the DOE grant DE-FG02-95ER40896, Outstanding  Junior Investigator Award, by the University of Wisconsin Research Committee
with funds provided by the Wisconsin Alumni Research Foundation, and
by the Alfred P.~Sloan Foundation.
  

\newpage

\end{document}